\documentclass[journal=jacsat,manuscript=article]{achemso}

\usepackage[version=3]{mhchem} % Formula subscripts using \ce{}

\usepackage{color}
\usepackage{amssymb}

\author{Vasily V. Temnov}
\affiliation[ITMO]{ITMO University, 197101 St. Petersburg, Russia}
\alsoaffiliation[IMMM_CNRS]{IMMM CNRS 6283, Le Mans Universit\'e, 72085 Le Mans, France} 
\email{vasily.temnov@univ-lemans.fr}
\author{Alexandr Alekhin}
\affiliation[IMMM_CNRS]{IMMM CNRS 6283, Le Mans Universit\'e, 72085 Le Mans, France}
\alsoaffiliation[CIC_nanoGUNE]{CIC nanoGUNE BRTA, E-20018 Donostia-San Sebastian, Spain}
\author{Andrei Samokhvalov}
\affiliation{ITMO University, 197101 St. Petersburg, Russia}
\author{Dmitry S. Ivanov}
\affiliation[ITMO]{ITMO University, 197101 St. Petersburg, Russia}
\alsoaffiliation[TU-KL]{Department of Physics and OPTIMAS Research Center, Technical University of Kaiserslautern, 67663 Kaiserslautern, Germany}
\author{Paolo Vavassori}
\affiliation[CIC_nanoGUNE]{CIC nanoGUNE BRTA, E-20018 Donostia-San Sebastian, Spain}
\alsoaffiliation[IKERBASQUE]{IKERBASQUE, Basque Foundation for Science, E-48013 Bilbao , Spain}
\author{Vadim P. Veiko}
\affiliation[ITMO]{ITMO University, 197101 St. Petersburg, Russia}

\title[An \textsf{achemso} demo]
{Nondestructive femtosecond laser lithography of Ni nanocavities by controlled thermo-mechanical spallation at the nanoscale}

\abbreviations{IR,NMR,UV}

\begin{document}

%%%%%%%%%%%%%%%%%%%%%%%%%%%%%%%%%%%%%%%%%%%%%%%%%%%%%%%%%%%%%%%%%%%%%
%% The "tocentry" environment can be used to create an entry for the
%% graphical table of contents. It is given here as some journals
%% require that it is printed as part of the abstract page. It will
%% be automatically moved as appropriate.
%%%%%%%%%%%%%%%%%%%%%%%%%%%%%%%%%%%%%%%%%%%%%%%%%%%%%%%%%%%%%%%%%%%%%
\begin{tocentry}

 \includegraphics[width=1.0\columnwidth]{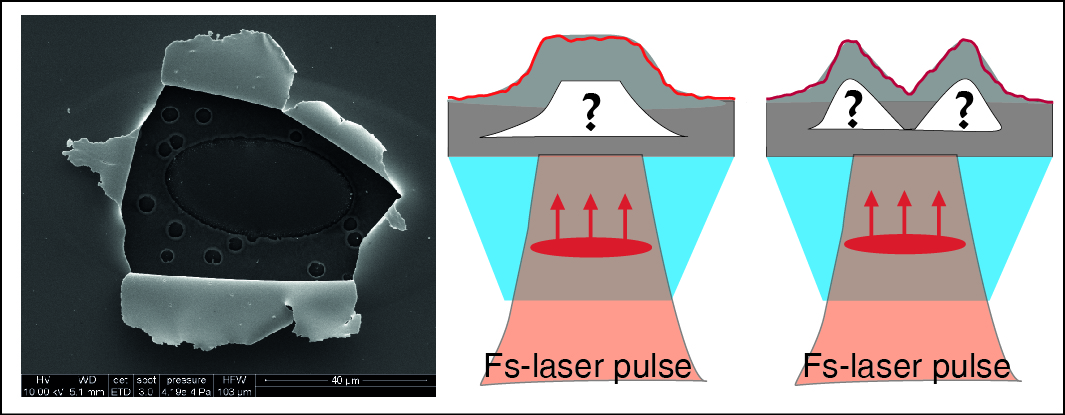}

\end{tocentry}

%%%%%%%%%%%%%%%%%%%%%%%%%%%%%%%%%%%%%%%%%%%%%%%%%%%%%%%%%%%%%%%%%%%%%
%% The abstract environment will automatically gobble the contents
%% if an abstract is not used by the target journal.
%%%%%%%%%%%%%%%%%%%%%%%%%%%%%%%%%%%%%%%%%%%%%%%%%%%%%%%%%%%%%%%%%%%%%
\begin{abstract}
We present a new approach to femtosecond direct laser writing lithography to pattern nanocavities in ferromagnetic thin films. To demonstrate the concept we irradiated 300~nm thin nickel films by single intense femtosecond laser pulses through the glass substrate and created complex surface landscapes at the nickel-air interface. Using a fluence above the ablation threshold the process is destructive and irradiation leads to the formation of 200~nm thin flakes of nickel around the ablation crater as seen by electron microscopy. By progressively lowering the peak laser fluence, slightly below the ablation threshold the formation of closed spallation cavities is demonstrated by interferometric microscopy. Systematic studies by electron and optical interferometric microscopies enabled us to gain an understanding of the thermo-mechanical mechanism leading to spallation at the solid-molten interface, a conclusion supported by molecular dynamics simulations. We achieved a control of the spallation process that enabled the fabrication of closed spallation nanocavities and their periodic arrangements. Due to their topology closed magnetic nanocavities can support unique couplings of multiple excitations (magnetic, optical, acoustic, spintronic). Thereby, they offer a unique physics playground, before unavailable, for magnetism, magneto-photonic and magneto-acoustic applications.  
\end{abstract}

%%%%%%%%%%%%%%%%%%%%%%%%%%%%%%%%%%%%%%%%%%%%%%%%%%%%%%%%%%%%%%%%%%%%%
%% Start the main part of the manuscript here.
%%%%%%%%%%%%%%%%%%%%%%%%%%%%%%%%%%%%%%%%%%%%%%%%%%%%%%%%%%%%%%%%%%%%%
\section{Introduction}

Destructive laser-matter interactions play an important role in the processing and structuring of materials. The direct optical excitation of solid surfaces and interfaces by single or multiple laser pulses
can be utilized to fabricate a variety of nanostructures~\cite{Vorobyev2013, Kim2007, Kim2009, Kuznetsov2009, Karstens2016, Jalil2019}. Nanostructuring occurs as a consequence of laser-induced phase transitions through a sequence of highly non-equilibrium states of matter \cite{Sokolowski1998, VonderLinde2000, Zhigilei2005} difficult to control.
Depending on laser and  material parameters, the laser-induced material destruction/modification mechanisms in absorbing solids can be either of thermal nature (melting, ablation, resoldification, photo-chemistry at surfaces and interfaces) or mechanical disruption (spallation) of the material, assisted by large amplitude acoustic/shock pulses. 

 In most cases the prerequisite of laser nanostructuring is laser-induced melting. Being an intricate phenomenon by itself, laser-induced melting triggers the complex spatio-temporal hydrodynamic and mechanical motion of melted material. Inevitable modification of resolidified surface landscapes may alter the material properties in an unpredictable manner preventing the use of fs-laser nanostructuring for applications in nanophotonics, where both nanometer precision and well-defined material properties are usually required at once.
 
 Our present investigation aims at circumventing the aforementioned complications and has been inspired by older experiments on macroscopically thick aluminum foils. Salzmann et al.~\cite{Salzmann1989} demonstrated that nanosecond optical excitation of the front side of 100~$\mu$m thick aluminum foil results in the formation of a closed cavity in the vicinity of its back side. The formation of the cavity has been explained by the local mechanical disruption (spallation) in solid aluminum induced by transient spatially inhomogeneous tensile stresses generated upon reflection of large amplitude compressive acoustic/shock pulses from the metal-air interface. Similar observations have been reported by Tamura et at.~\cite{Tamura2001} using picosecond and femtosecond laser pulses on free-standing aluminum foils. However, owing to the macroscopic thickness of the investigated aluminum layers, these studies were not topical from the perspective of modern nanotechnologies.  

Technically similar experimental investigations on metallic thin films with the thicknesses in the range between 10 and 400~nm suggested that fs-laser pulses result either in the removal of the entire film from the substrate and  the formation of an ablation crater (punching~\cite{Domke2014}) or to a local thermoelastic separation of the entire film from the dielectric substrate,~\cite{Meshcheryakov2005} sometimes denoted as \textquotedblleft blistering\textquotedblright~\cite{Inogamov2015} or \textquotedblleft bulging\textquotedblright~\cite{Domke2014}. 

In this paper we introduce a new way to achieve a deterministic femtosecond laser nanostructuring {\it inside} a 300~nm nickel thin film which is partially melted by a single femtosecond laser pulse focused through a glass substrate. The concurrent fast resolidification of the melted material and the spallation dynamics, lead to the formation of a dome-like cavity entirely made by nickel and probably enclosing vacuum. Being neither bulging nor punching, this new mechanism of thermo-mechanical spallation manifests itself in the deterministic, i.e. controllable, separation of a 200~nm layer of solid nickel (ceiling of the dome) from a 100~nm layer of laser-melted nickel remaining on the glass substrate (floor of the dome). The deterministic formation of closed spallation cavities with dimensions that can reach the sub-micron range as well as their periodic arrangements are quantified by optical interferometric microscopy \cite{Temnov2006}. Such periodic landscapes open the door for numerous technological applications of this novel and potentially disruptive fs-laser-based nanostructuring technology. From the one side, it allows to approach fundamentals questions of modern science related to the physical properties of novel states of matter, such as amorphous spin glasses~\cite{Zhong2014} or curved magnets~\cite{Streubel2016}. From the other side, such magnetic nanocavities can support elementary excitations of different character (magnetic, optical, acoustic...), thereby providing fundamental conditions for studying and exploiting  their coupling.

We have chosen nickel thin films to demonstrate the potential of the proposed approach for several reasons.  First of all, functional nanostructuring of nickel has a great relevance for applications since studies of elementary excitations in nickel thin films and nanostructures revealed already a tremendous variety of interesting fundamental physical phenomena such as ultrafast laser demagnetization~\cite{Beaurepaire1996}, optical and acoustic generation of exchange magnons~\cite{VanKampen2002, Besse2020}, magneto-acoustic driving of ferromagnetic resonance ~\cite{Weiler2011, Kim2012, Janusonis2016, Chang2017}, magneto-elastic switching~\cite{Thevenard2013, Vlasov2020}, magnetic effects on the photonic properties~\cite{Tran2018, Belotelov2019, Maccaferri2020}, the extraordinary optical transmission through periodic arrays of sub-wavelength holes~\cite{Ebbesen1998, Coe2008}, etc. Second, reliable input parameters for molecular dynamics simulations of fs-laser-excited nickel are available \cite{Ivanov2003}, thereby facilitating a deeper understanding of the involved phenomena. Third, the physics of nickel-based complex magnetic compounds is particularly exciting since it is spanning from unique magnetic properties of 3d-metal alloys such as as Permalloy ($Ni_{80}Fe_{20}$, Py)~\cite{Elmen1936} to ultrafast magnon spectroscopy dynamics of rear-earth doped Py~\cite{Salikhov2019} to the high-$T_{C}$ superconductivity in nickelates~\cite{Li2019}. An experimental technique for shaping thin films of complex magnetic compounds at the nanoscale without destroying their unique physical properties could revolutionize the respective fields of research.  

\section{Microscopic characterization of nickel thin films irradiated through glass substrate}

\begin{figure}
    \includegraphics[width=0.8\columnwidth]{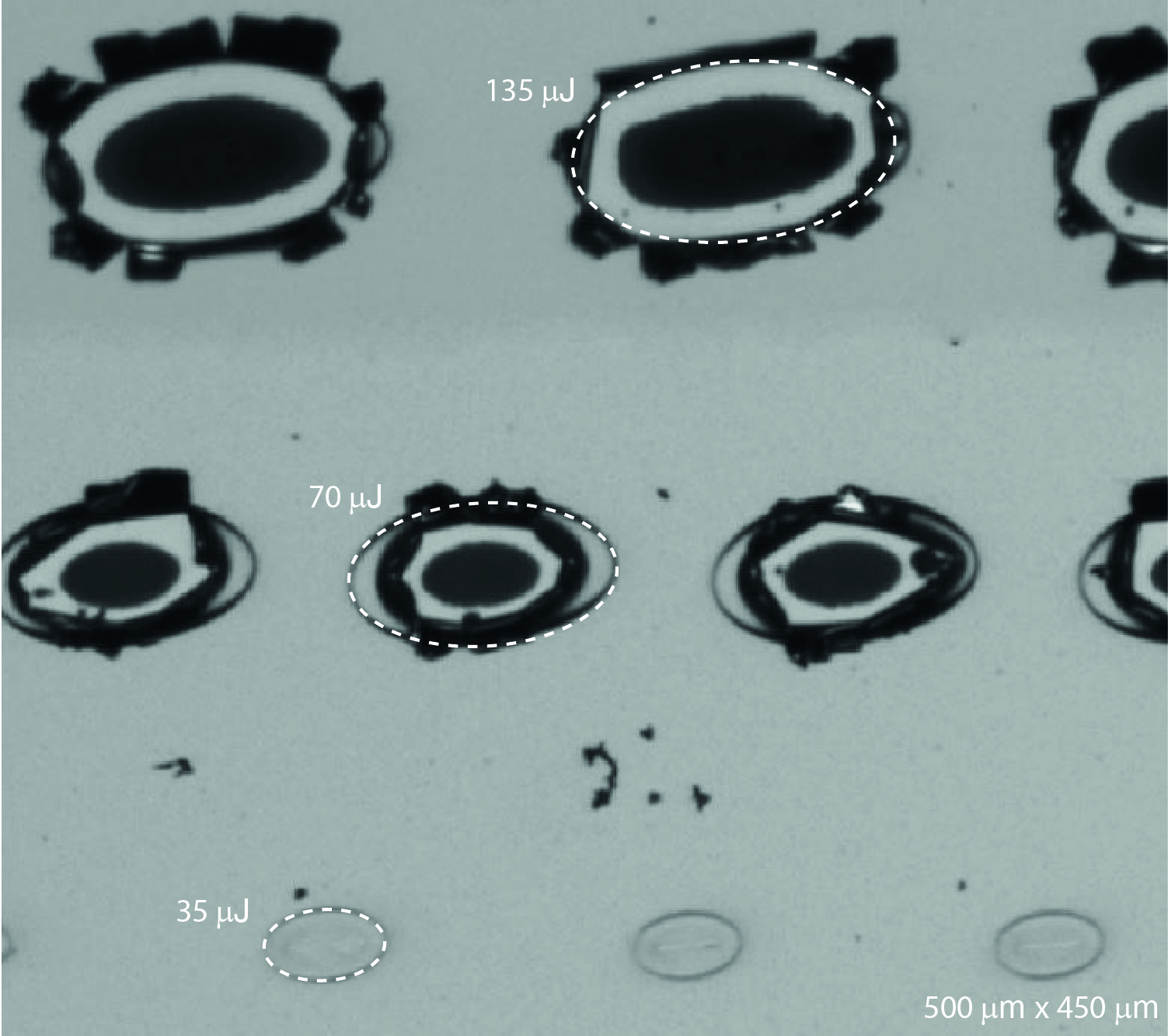}
  \caption{(Color online) Reflection optical microscopy of laser damage spots on a 300~nm thin 
  nickel film on glass irradiated with single femtosecond laser pulses focused through 
  the glass substrate. The sample was moving horizontally at different speeds
  to ensure interaction of pulses from a 10 Hz Ti:Sa pulse train at fresh spots on the 
  surface. Pulse energies were 135, 70 and 35~$\mu$J, respectively. 
  Complex pulse energy-dependent damage patterns signify a new physical phenomenon:  
  fs-laser-induced spallation of partially melted metallic thin film.}
  \label{fgr:300nm_Ni_optmicro}
\end{figure}

During the first set of experiments, the train of ultrashort laser pulses from an amplified Ti:Sa laser (100~fs duration, 10~Hz repetition rate, pulse energy up to 140~$\mu$J) was focused with a quartz lens (focal length 40~cm) under 45 degrees on a 1~cm$x$1~cm sample, scanning the beam horizontally at a constant speed thereby making possible to fabricate sequences of almost identical structures in the single-shot regime. Thanks to the high reliability of such scanning fs-laser patterning, we then achieved the fabrication of periodic arrays of different shapes and periods. We first started with the investigation of the key parameters that need to be controlled to achieve the desired precision and fidelity of the process. Figure~\ref{fgr:300nm_Ni_optmicro} presents an optical image of the surface of a 300-nm thick Ni film irradiated with single femtosecond laser pulses at different spots through fused silica substrate. Here, one can observe three horizontal lines of structures obtained with different pulse energies of 35, 70 and 135~$\mu$J, respectively. During the irradiation, the sample was moved horizontally in order to ensure that each structure is produced by a single laser shot at a fresh spot. If the pulse energy is sufficiently large (70 and 135~$\mu$J in Fig.~1), its absorption results in the formation of rather complex damaged areas. Dark elliptical areas represent ablation craters where the entire Ni-film is removed from the surface: the low reflectivity is due to reflection from the glass-air interface at the bottom of the crater. The elliptical ring area just outside the ablation crater displays the same optical reflectivity of nickel. It is surrounded by irregularly-shaped dark areas, which are identified as quasi-freestanding flakes of nickel film. A close inspection of Fig.~\ref{fgr:300nm_Ni_optmicro}, evidences the existence of an additional outer ring-like elliptical area, which becomes the entire elliptical spot in the case of the weaker pulse energy of 35~$\mu$J. The weak optical contrast displayed by this region suggests that the film hasn't been destroyed but rather structurally modified, i.e. lifted from the substrate. This conclusion is confirmed by the results of optical interferometric microscopy of these structures (to be discussed below), which allow us to state that in this area we are above the threshold for the thermo-mechanical spallation process inside the film. We determined both ablation and spallation thresholds using the Liu method~\cite{Liu1982} assuming the Gaussian distribution of laser fluence. For a 300~nm thick nickel films excited through the substrate the ablation and spallation thresholds were found to be  $F_{\mbox{abl}}=(950\pm100)~\mbox{mJ}/\mbox{cm}^2$ and $F_{\mbox{spall}}=(480\pm 50)~\mbox{mJ}/\mbox{cm}^2$, respectively.

\begin{figure}
    \includegraphics[width=0.8\columnwidth]{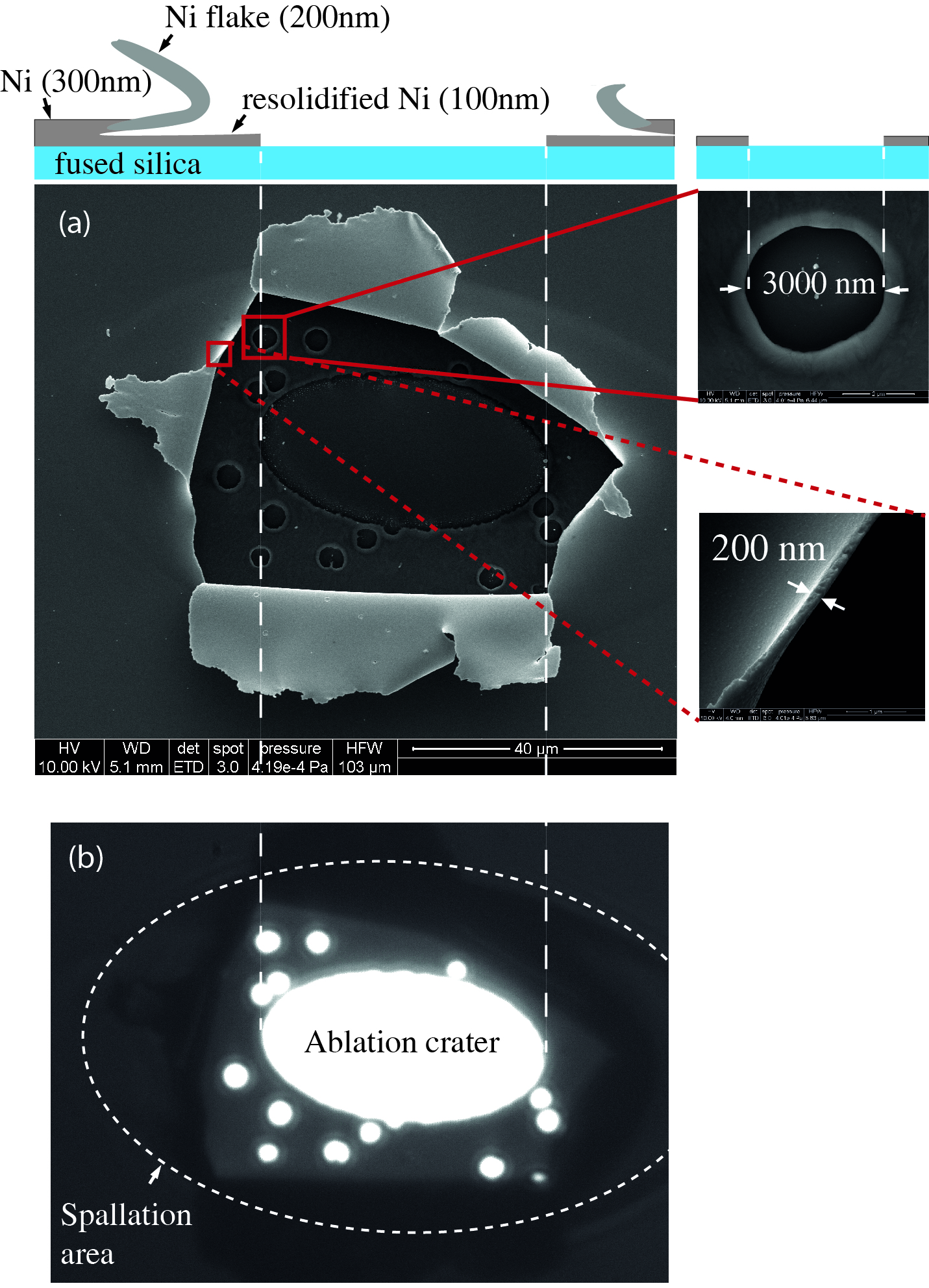}\\
  \caption{(Color online) (a) Scanning electron microscopy demonstrates that the laser-damaged area
consists of an elliptical hole (ablation crater) in the center of laser-excited area surrounded 
by random-shape free-standing nickel flakes of 200~nm thickness. The part of the film remaining on the glass surface contains round holes with a diameter of approximately 3 micrometers. (b) Optical transmission microscopy confirms the removal of the entire film in the elliptical ablation crater and round holes.}
  \label{fgr:300nm_Ni_SEM}
\end{figure}

Scanning electron microscopy (SEM) of a typical damage area induced by at 70~$\mu$J laser pulse in Fig.~\ref{fgr:300nm_Ni_SEM} provides a detailed microscopic view of uncontrolled spallation, which is nevertheless useful to understand key aspects of the process. A complementary optical transmission microscopy on the same structure helps to visualize multiple areas where the film has been removed from the surface: an elliptical ablation crater and micrometer-small round holes outside it. 

Outside the elliptical ablation crater the laser fluence decreases towards the periphery of the crater; at the certain distance from the center the absorbed energy is not sufficient to remove the Ni film completely. Nevertheless, in this region the local laser fluence is sufficient to melt a fraction of the film interfaced with the substrate and lift up and rip the topmost part. This outcome is the result of an uncontrolled spallation as confirmed by the molecular dynamics simulation discussed in the last section. 
As a result of this thermo-mechanical spallation mechanism, the melted part of the Ni film remains on the substrate and the solid part forms dangling flakes surrounding the crater. According to the SEM images presented in Figure~\ref{fgr:300nm_Ni_SEM}, the thickness of the dangling flakes is about 200 nm. In both optical and SEM images in Fig.~\ref{fgr:300nm_Ni_optmicro} and~\ref{fgr:300nm_Ni_SEM}, it is interesting to note round micrometer-size holes outside the ablation crater. The part of the Ni film interfaced with the substrate is melted and remains in a molten and overheated state for nanoseconds. The complex dynamics of phase transitions in laser-excited Ni caused the appearance of such round holes. The formation of these round holes, as well as their number and size, in the outer part of the crater depends on the kinetics of solid-liquid (melting) and the subsequent liquid-solid (resolidification) phase transition of the metal film remaining on the substrate. Optical microscopy on a fs-laser irradiated Ni/sapphire sample did not evidence any holes outside the ablation crater, presumably due to the much faster cooling and resolidification rates of laser-melted nickel on sapphire substrate characterized by the much larger thermal conductivity as compared to fused silica.

A close inspection of Figure~\ref{fgr:300nm_Ni_optmicro}) shows that the region of laser-produced structures progressively shrinks up on reducing the pulse energy, below a certain threshold (between 36 and 70~$\mu$J in the case of the 300~nm thick Ni film studied here), only an elliptical region of controlled spallation (i.e. without the formation of ablation craters and dangling Ni flakes) remains: compare the transition from the first to the third line of structures in Fig.~\ref{fgr:300nm_Ni_optmicro}).

%-------------
%-------------

\section{Formation of closed spallation cavities below the ablation threshold}

In the following we demonstrate that controlled spallation can be indeed achieved tuning the pulse energy in the range between spallation and abalation onsets.  Figure 3 shows the interferograms of nickel sample excited below the ablation threshold and slightly above the spallation onset, namely playing with the laser pulse energy in the neighbourhood of 35~$\mu$J. Interferograms from the front side display pronounced shift of optical interference fringes. Comparison with reference interferogram of unexcited surface can be used to reconstruct the phase landscape $\Delta\Psi(x,y)$ by applying, for example, the 2D-Fourier transform algorithm~\cite{Temnov2006}. For fluences below the ablation threshold the optical properties of the nickel-air interface are not affected by laser-induced  phase transitions. Therefore, the phase shift can used to unambiguously determine the surface displacement $d(x,y)=(\lambda/4\pi)\Delta\Psi(x,y)$, where $\lambda$=650~nm is optical wavelength used for interferometric imaging.  

\begin{figure}
    \includegraphics[width=0.8\columnwidth]{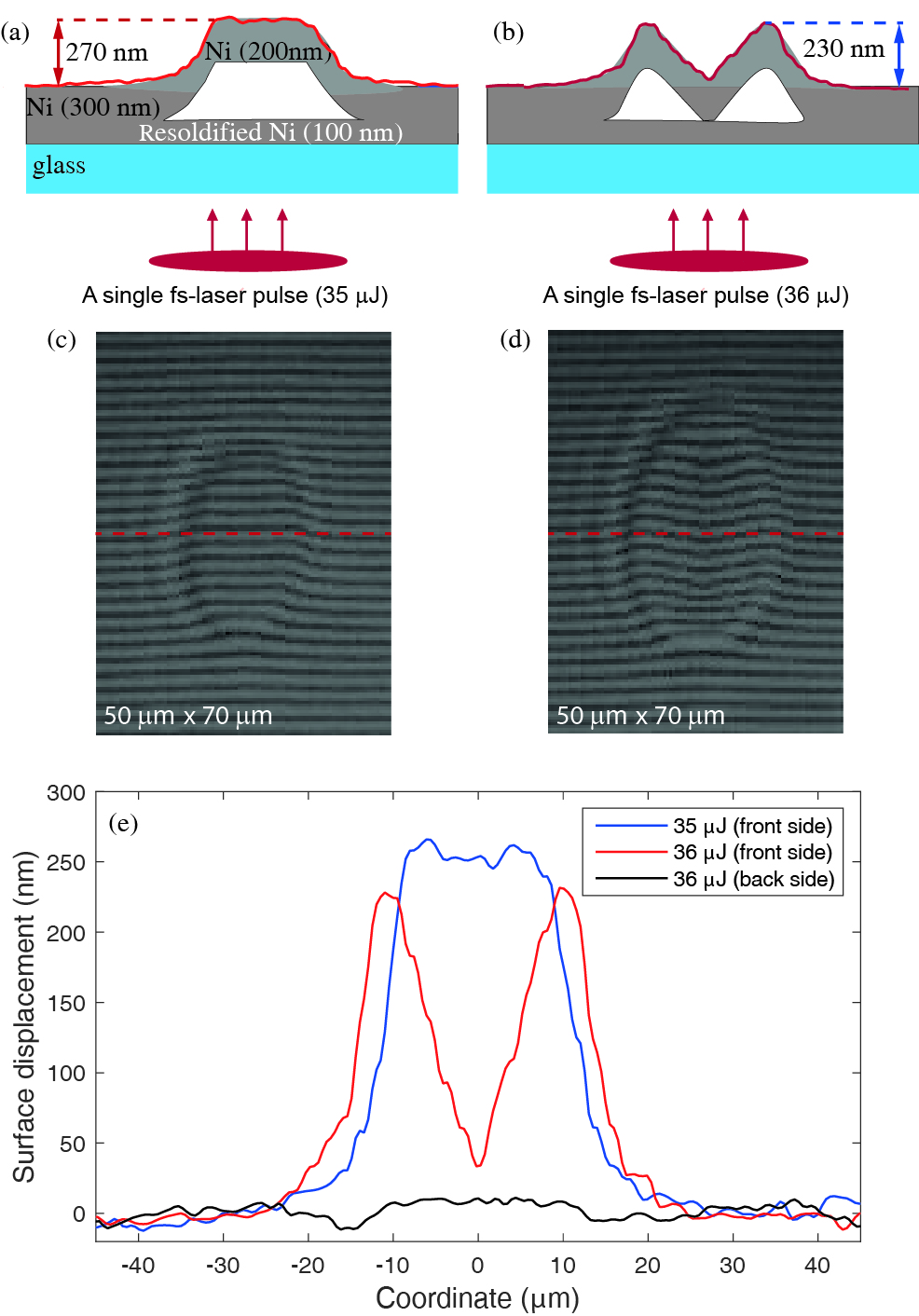}\\
  \caption{(Color online) Interferometric microscopy of Ni-film irradiated at low laser fluences of $F=1.17~F_{spall}$ (pulse energy $35~\mu\mbox{J}$, (c)) and $F=1.2~F_{spall}$ (pulse energy 36~$\mu$J, (d)) evidence distinct surface displacements with maximum amplitudes of (a) 270~nm and (b) 230~nm, respectively. The lack of interferomeric phase shifts within laser-irradiated spots from the glass side  (not shown) and the results of Fig.~\ref{fgr:300nm_Ni_SEM}(a) make it possible to reconstruct the distinct {\it flat-top} and {\it M-shaped} internal structures of closed fs-laser-produced spallation cavities. (e) Surface profiles along the dashed horizontal lines in (c) and (d) obtained from the Fourier-reconstructed interferometric phase maps~\cite{Temnov2006}.}
  \label{fgr:Interferometry}
\end{figure}

Laser pulses with energy 35~$\mu$J (peak fluence $F=1.17F_{spall}$) result in a {\it flat-top} surface profile (Fig.~\ref{fgr:Interferometry}(a,c)), where a spalled 200~nm-thin solid layer of nickel lifted up to a height of 270~nm from the surface, without ripping the film. Indeed, the preserved continuity of the film is the mechanism that decelerates and eventually stops the lifting up of the spalled layer thanks to the mechanical forces exerted by the un-irradiated film on the sides,resulting in the formation of a {\it flat-top} surface profile. Excitation with a slightly higher energy of 36~$\mu$J (peak fluence $F=1.20F_{spall}$) generated a characteristic {\it M-shaped} surface profile (Fig.~\ref{fgr:Interferometry}(b,d)). In this case the mechanical forces were sufficiently strong to induce the plastic deformation of the spalled nickel film that bounced back towards the surface in the center of the laser-excited spot. The outer part of the spallation cavity is elevated by 230~nm on top of the undisturbed nickel surface. 
Control interferograms from the back side of the sample (not shown) displayed only minor interferometric phase shifts, which could be explained by the change in the optical constants of the resolidified nickel. Therefore we conclude that part of molten nickel resolidifies on the glass surface leading to the peculiar spallation cavities sketched in the cross sections shown in Fig.~\ref{fgr:Interferometry}(a,b,e) for 35 and 36~$\mu$J pulse energies, respectively.

%-------------
%-------------

One of the most intriguing questions is what is inside the spallation cavities. If gas diffusion through the Ni film could be ruled out, the volume enclosed by the cavity should be in vacuum conditions and the estimated vacuum pressure would be given by the saturated vapor pressure of nickel at the room temperature. In this case, the vaccuum pressure would not exceed $10^{-26}$~Pa~\cite{Alcock1984, Haynes2016}, thereby suggesting the formation of a cavity enclosing space in ultra-high vacuum conditions. However, the actual nickel films are polycrystalline so that permeation of different gases from air is expected to occur after a some time resulting in a certain life-time of the vacuum conditions. Whereas permeation of oxygen through nickel is rather inefficient at room temperature~\cite{Park1987}, the atomic hydrogen possesses the higher permeation rate ~\cite{Furuya1984}. Being not present under ambient conditions, the permeation of atomic hydrogen through nickel has been evidenced  under ambient conditions~\cite{Phillips1964}. Estimation of the diffusion of hydrogen atoms through polycrystalline nickel~\cite{Furuya1984} for a 200~nm nickel foil of the observed spallation cavities provide the diffusion time of the order of 100~$\mu$s, which represents the lower limit of vacuum life time in fs-laser-produced spallation cavities.   

Therefore we speculate that the transient vacuum inside the spallation cavity protects the process of resolidification of the bottom layer from surface chemistry, which would alter the Ni surface material under ambient conditions and elevated temperatures~\cite{Furuya1984}. 
Cooling down and resolidification of laser-melted nickel layer on glass substrate is governed by thermal diffusion in glass. The cooling rate of a thin liquid layer on a dielectric substrate is determined by its thickness, the thermal boundary resistance of liquid-substrate interface and the heat conductivity of the substrate~\cite{Cahill2002}. We estimate that this process takes place on the nanosecond time scale in the case studied here. The thickness of fs-laser-melted layer could be preselected by adjusting the laser pulse energy, whereas the cooling rate can be engineered via a proper choice of the substrate. Using substrates with a high thermal conductivity can produce ultrahigh cooling rates of the order 10$^{13}$~K/s, under which amorphization of monoatomic liquid metals become possible~\cite{Zhong2014}. Given the case that our conclusions about the transient vacuum are correct, our experiments may represent an alternative way for creating thin films of high-purity amorphous metals under ambient conditions.     

%-------------
  \begin{figure}
    \includegraphics[width=0.8\columnwidth]{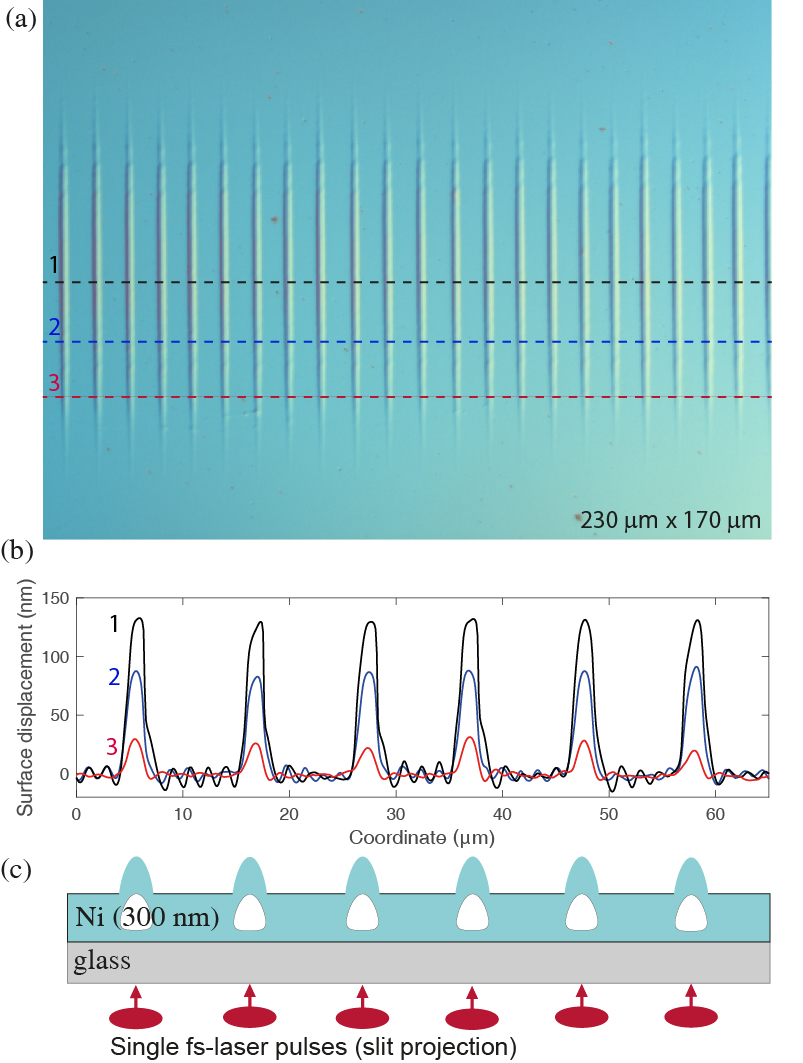}\\
  \caption{(Color online) (a) Differential interference microscopy of a periodic array of elongated spallation cavities in a 300~nm Ni film produced by moving the sample through the line focus (see text for details). (b) Interferometric microscopy evidences regular periodic surface excursion consisting of narrow bell-shaped bumps with amplitudes reaching 130~nm. (c) The internal structure of a periodic landscape of closed spallation cavities providing surface excursions in (b).}
  \label{fgr:Interferometry2}
\end{figure}
%------------

Using the slit projection technique~\cite{Kuznetsov2009}, we have produced periodic landscapes of elongated spallation cavities. A train of unfocused femtosecond pump pulses (pulse energies 35-140~$\mu$J, repetition rate 10~Hz) illuminated a 150~$\mu$m wide slit, which was projected with a microscope objective on the nickel-glass interface approximately $3\times70$~$\mu$m line focus. The sample was moving at a constant speed of 0.1~mm/s through the line focus to produce an array of identical structures with 10~$\mu$m period. Differential interference contrast microscopy in Fig.~\ref{fgr:Interferometry2}(a) shows an image of narrow lines. Imaging optical interferometry (not shown) provide surface profiles in Fig.~\ref{fgr:Interferometry2}(b): narrow bell-shaped bumps with spatial width below $2~\mu\mbox{m}$ and reaching a height of 130~nm in the center (cross-section 1) and gradually decreasing to zero at the edges of the cavities (cross-sections 2 and 3). These results demonstrate a pathway to fast nanofabrication of cavities with different shape in one shot. By changing by the velocity of the sample from 0.3 to 0.03~mm/s we were able to produce gratings with periods from 30 to 3 $\mu$m. However, at small velocities the control of mechanical vibrations becomes more difficult and this affects the long-time stability of the laser beam focusing, becoming a potential issue for the production of periodic diffraction grating with large pitch. 

\section{Molecular dynamics simulations}
In order to deeply understand the mechanism of the spallation process observed in the experiments, we performed the atomistic-continuum modeling of fs-laser induced dynamics for Ni thin films irradiated through glass substrate. The model is based on the Molecular Dynamics (MD) approach, but, due to integrated Two Temperature Model (TTM)~\cite{Ivanov2003, Zhigilei2004}, it also accounts for the diffusion of free carriers in metals, playing a determinant role in case of ultrashort laser pulse interactions. The combined MD-TTM model provides a detailed atomic-level description of the laser light absorption by conduction band electrons, the energy transfer to the lattice due to electron-phonon coupling and electron heat diffusion, resulting in the kinetics of multiple fast non-equilibrium laser-induced defragmentation (spallation) processes. The strong and ultrafast heating/melting results in a high internal stress of 20~GPa developed in the vicinity of the substrate within the first 5 ps that launch a compression wave propagating throughout the film, with a typical amplitude of several GPa. Upon reflection from the Ni-air interface this pressure wave is converted in a tensile pulse, generating a reflected tensile wave with amplitude of about 4~GPa that reaches the solid-liquid interface, where it triggers the spallation process by creating voids in the laser-melted layer remaining on the surface. The results of MD-TTM simulations depend critically on the applied laser fluence, the acoustic mismatch and mechanical adhesion at the Ni-glass interface. At a given fluence, the latter property decides whether the laser-melted layers remains on the surface or is removed, forming the ablation crater. Quantitative understanding of this phenomenon requires substantial computational efforts but also experiments on Ni thin films of different thickness, both beyond the scope of this paper. Nevertheless the MD-TTM simulations confirm that the experimentally observed 200~nm layer of Nickel film is spalled in the vicinity of the solid-liquid interface. The observed holes in the remaining 100~nm layer of resolidified nickel fingerprint the formation of voids in liquid nickel by transient negative pressures.        

\section{Conclusions and outlook}
We demonstrated that the controlled thermo-mechanical spallation of nickel films can be induced and controlled by irradiation with a single femtosecond laser pulses. Molecular dynamics simulations shade light at the complex physics of the spallation process occuring in inside the nickel film at an interface between solid and (laser-molten) nickel. By tuning the irradiation parameters, closed spallation cavities with two distinct flat and "M" shaped top have been produced and characterized by high-resolution interferometric microscopy. Using slit projection technique periodic arrays of elongated bell-shaped spallation cavities with periodicities in the range 3 - 30~$\mu$m and heights up to 200~nm have been produced demonstrating that the process can be finely controlled to induce desired shapes and parallelized for the fast fabrication of metasurfaces and gratings for visible and IR, potentially also for THz, radiation. The flexible nature of the quasi-free-standing outer nickel membrane forming the spallation cavities  will allow for substantial and fast modulation of light diffraction by exciting the acoustic eigenmodes of the cavities. The curved nature of the nickel membrane is expected to result in unique topology-induced magnetic states, which are the focus of intense studies~\cite{Streubel2016}. More in general, such magnetic nanocavities can support a variety of excitations: from magnetic to optical and acoustic ones, thereby providing a unique multifunctional metasurface for studying their properties and the physics arising from their coupling. 
Our results are encouraging and indicate that such back-side femtosecond laser lithography can be further advanced through the use of (i) more complex spatial patterns of pump radiation to create sophisticated surface landscapes  and (ii) magnetic materials with tailored physical properties. We envision that future investigations of acoustic, magnetic, magnetophotonic, magneto-acoustic and 'vacuum' properties of spallation cavities could open a new avenue in modern nanoscience.

%%%%%%%%%%%%%%%%%%%%%%%%%%%%%%%%%%%%%%%%%%%%%%%%%%%%%%%%%%%%%%%%%%%%%
%% The "Acknowledgement" section can be given in all manuscript
%% classes.  This should be given within the "acknowledgement"
%% environment, which will make the correct section or running title.
%%%%%%%%%%%%%%%%%%%%%%%%%%%%%%%%%%%%%%%%%%%%%%%%%%%%%%%%%%%%%%%%%%%%%
\begin{acknowledgement}

The authors greatly acknowledge Michael Farle from Duisburg-Essen university for stimulating discussions.  Funding through the ITMO Fellowship and Professorship program, Strategie internationale “NNN-Telecom” de la Region Pays de La Loire, the Spanish Ministry of Economy, Industry and Competitiveness under the Maria de Maeztu Units of Excellence Programme (MDM-2016-0618), and from the European Union under the Project H2020 FETOPEN-2016-2017 “FEMTOTERABYTE” (Project n. 737093) is gratefully acknowledged.\ldots

%{\color{green}Please use ``The authors thank \ldots'' rather than ``The authors would like to thank \ldots''.

%The author thanks Mats Dahlgren for version one of \textsf{achemso}, and Donald Arseneau for the code taken from \textsf{cite} to move citations after punctuation. Many users have provided feedback on the class, which is reflected in all of the different demonstrations shown in this document.}

\end{acknowledgement}

%%%%%%%%%%%%%%%%%%%%%%%%%%%%%%%%%%%%%%%%%%%%%%%%%%%%%%%%%%%%%%%%%%%%%
%% The same is true for Supporting Information, which should use the
%% suppinfo environment.
%%%%%%%%%%%%%%%%%%%%%%%%%%%%%%%%%%%%%%%%%%%%%%%%%%%%%%%%%%%%%%%%%%%%%
%\begin{suppinfo}

%This will usually read something like: ``Experimental procedures and characterization data for all new compounds. The class will automatically add a sentence pointing to the information on-line:

%\end{suppinfo}

%%%%%%%%%%%%%%%%%%%%%%%%%%%%%%%%%%%%%%%%%%%%%%%%%%%%%%%%%%%%%%%%%%%%%
%% The appropriate \bibliography command should be placed here.
%% Notice that the class file automatically sets \bibliographystyle
%% and also names the section correctly.
%%%%%%%%%%%%%%%%%%%%%%%%%%%%%%%%%%%%%%%%%%%%%%%%%%%%%%%%%%%%%%%%%%%%%
%\bibliography{achemso-demo}

\providecommand{\latin}[1]{#1}
\makeatletter
\providecommand{\doi}
  {\begingroup\let\do\@makeother\dospecials
  \catcode`\{=1 \catcode`\}=2 \doi@aux}
\providecommand{\doi@aux}[1]{\endgroup\texttt{#1}}
\makeatother
\providecommand*\mcitethebibliography{\thebibliography}
\csname @ifundefined\endcsname{endmcitethebibliography}
  {\let\endmcitethebibliography\endthebibliography}{}

\end{document}